\documentclass[aps,prl,amsmath,twocolumn]{revtex4}
  \usepackage{bm}
\usepackage{graphicx}
 \usepackage{amssymb}

\def \bomega{\boldsymbol{\omega}}

\begin{document}

\title{Longitudinal and transverse velocity scaling exponents from
merging of  the Vortex filament and  Multifractal models. }
\author{K. P. Zybin\footnote{Electronic address: zybin@lpi.ru} }   
\author{V. A. Sirota}
\affiliation{P.N.Lebedev Physical Institute of RAS, 119991, Leninskij pr.53, Moscow,
Russia }
%

\begin{abstract}
We suggest a simple explanation of the difference between
transverse and longitudinal scaling exponents observed in
experiments and simulations. Based on the Vortex filament model
and Multifractal conjecture, we calculate both scaling exponents
for any $n$ without any fitting parameters and ESS anzatz. The results  are in
very good agreement with the data of simulations.

\end{abstract}

\maketitle

Numerous direct numerical simulations (DNS) of hydrodynamical
turbulent flow, as well as high-resolution experiments, have been
performed during last twenty years \cite{25years}. 
The  positive result of these investigations is that 
numerical and experimental
approaches demonstrate a very good agreement. In particular,
velocity statistics calculated from DNS  practically coincides
with that obtained from experiments
\cite{25years}.  This justifies 
 the accuracy of both kinds of results. But there
still is a lack of physical understanding of the processes
occuring in a turbulent flow and contributing to statistics. Which
of the infinite number of solutions to the NSE are responsible for
the observed intermittent behavior, and why?

Okamoto et al. \cite{FargeOkamoto}
 claimed that  the most part of  the flow's helicity and,
in some sense, the most part of information about the flow are
contained in the same small part of the liquid volume spread more
or less uniformly throughout the liquid. These structures are
stable, their lifetime exceeds many times the largest eddy
turnover time in the flow.

This seems us to be a very important step.
Probably, these elongated regions of high vorticity are just the 
structures that determine the small-scale
statistics of a turbulent flow. 
Thus, it would be useful to elaborate some physical understanding of these objects.

Such an attempt was performed in 
\cite{JETP1,PRL1,JETP2,PRL2,Scripta,PRE,PhysicaD,PRE2}.
The main idea is that  in the regions where vorticity is high (vortex filaments), the
vorticity  itself stabilizes the motion. So, despite the
stochastic large-scale forces that give energy to a filament, the
motion inside the filament has an essential non-stochastic
component. This may be the reason of stability  of filaments. The
random large-scale forces then act, on average, as a stretching
force. Stretching (not breaking!) of vortices is the mechanism
that provides the observed energy flux from larger to smaller
scales, and observed statistical properties of turbulence. This is
the main difference of our approach from the Kolmogorov's approach
developed in the K41 theory \cite{Frisch}.

The relation between longitudinal and transversal Euler velocity structure functions
 is one of the problems that have had no answer up to now. These functions are
defined by
$$  
S_n^{\parallel}(l)=\left< \left| \Delta {\bf v}\cdot \frac{\bf l}{l}\right| ^n\right>
\ , \quad
S_n^{\perp}(l)=\left< \left| \Delta {\bf v} \times \frac{\bf l}{l}\right|
^n\right>
$$  
Here
$\Delta {\bf v} = {\bf v}({\bf r}+{\bf l}) -{\bf v}({\bf r}) $
is velocity difference between two near points separated by ${\bf l}$,
 and the average is taken over all pairs of points separated by given $l$.

 Experiments and DNS show that inside the inertial range the structure functions obey scaling laws,
  $$ S_n^{\perp}(l)\propto l^{\zeta_n^{\perp}} \ , \quad S_n^{\parallel}(l)\propto l^{\zeta_n^{\parallel}} \ .$$
  The scaling exponents $\zeta_n^{\perp}$, $\zeta_n^{\parallel}$  are believed to be independent on conditions of the experiment. They are intermittent, i.e., $\zeta_n^{\perp}/n$ and $\zeta_n^{\parallel}/n$ are decreasing functions of~$n$.

  The question whether the two scaling exponents coincide in isotropic and homogeneous turbulence
  or not, is open. There is an exact statement that $\zeta_n^{\parallel}=\zeta_n^{\perp}$  for $n=2$ and 3 \cite{Frisch}.
 On the other hand, modern
experiments \cite{Zhou,Shen} and numerical simulations
\cite{Chen,Gotoh,Benzi,Arn} show some
significant difference between $\zeta_n^{\parallel}$ and $\zeta_n^{\perp}$ at higher
$n$. But the proponents
of the equality argue that the difference may result from finite Reynolds
number effects \cite{He,Hill,L'vov97} or anisotropy
\cite{BifProc}.

The aim of the paper is to analyze the question by means of the Vortex filament model.
We show that the divergence of the scaling exponents can be understood and calculated from 
general physical considerations 
using the tools provided by the Vortex filament model and the well-known Multifractal (MF) conjecture.

Before we consider  the problem from the viewpoint of the Vortex
filament model, we note that isotropic and homogeneous medium
does not necessarily imply $\zeta_n^{\perp}=\zeta_n^{\parallel}$. For a simple illustration,
consider a gas of tops, each rotating around its own axis, the
axes directions  distributed randomly. Although locally a strong
asymmetry could be found near each top, the whole picture remains
isotropic. The Vortex filament model provides analogous situation that might be in real turbulence.

In  \cite{PRE2} we suggested to join our model of vortex filaments
with the MF model to get velocity scaling
exponents. We now remind briefly some ideas of the two models and
the results of \cite{PRE2}. Then we proceed to the difference
between the longitudinal and transverse scaling exponents.

\vspace{0.5cm}

The MF model implies that  the determinative contribution to
velocity structure functions  is given by the solutions (regions)
where
$$
\Delta v(l) = | {\bf v }({\bf r} + {\bf l}) -{\bf v}({\bf r}) |   \sim l^{h ({\bf r}) }
$$
So, to calculate structure functions, it is enough to consider only a set  of scaling solutions that can be numbered  by $h$.
This property is called local scale invariance.

The Large Fluctuations Theorem states that the probability of measured velocity difference
$\Delta v (l)$ to have the scaling $h$ is a power-law function of $l$:
$$  \texttt{P} \propto l^{3-D(h)} $$
Knowing $D(h)$, one could in principle calculate all structure functions:
$$
\langle \Delta v^n \rangle =
\int \delta v^n \texttt{P} dh =
 \int l^{nh} l^{3-D(h)} d\mu(h)
$$
Here $d\mu(h)$ is the measure that is responsible for relative weights of different values $h$.
In the limit
$l\to 0$, only the smallest exponent contributes to the integral;
it then follows
\begin{equation}
\label{MF}
\zeta_n= \min \limits_h \left( nh+3-D(h)\right)
\end{equation}
%
Without loss of generality, one can treat $D(h)$ as  a concave function.

As follows from (\ref{MF}),
the point $h_0$ where
$D(h)$ reaches its maximum corresponds to $n=0$: the requirement
$\zeta_0=0$ leads to $D(h_0)=3$.

The definitional domain of $D$ is restricted by some $h_{min}$ and $h_{max}$. Since
we are interested in positive values $n$,
we will hereafter restrict ourselves by $h\le h_0$. What about $h_{min}$, from  the condition  \cite{Frisch}  $\zeta''_n<0$ and the finiteness of the Mach number (or from the absence of singularities of $\Delta v $ as $l\to 0$) it follows $ \zeta'_n \ge 0$ for any $n$.
The minimum in (\ref{MF}) is reached at $h_n=\zeta'_n$. Since negative values of the derivative
are forbidden, we get  $h_{min}\ge 0$. The behavior of $D(h)$
near $h_{min}$  determines the behavior of $\zeta_n$ at
$$
n\simeq n_* = D'(h_{min}) \ .
$$
For larger $n$, the minimum in  (\ref{MF}) is reached at the boundary $h=h_{min}$. Thus,
the behavior of $\zeta_n$ at very large $n$ depends only on the value of $h_{min}$:
 the asymptotic behavior of $\zeta_n$ for all non-zero $h_{min}$  is
\begin{equation}
\label{greatern}
\left.  \zeta_n \right|_{n>n_*} =  n h_{min} +3- D(h_{min}) \sim _{n\to \infty} n h_{min}
\end{equation}
If $h_{min}=0$, the possible growth of $\zeta_n$ at $n>n_*$ is steeper than linear; if $D(h_{min})$ is finite, we get $\left. \zeta_n \right| _{n>n_*} = const $.

\vspace{0.5cm}

We now proceed to the Vortex filament model.
In \cite{JETP1} we derived the equation
\begin{equation}\label{main}
\ddot\omega_i = \rho_{ij}\omega_j \ , \quad  \rho_{ij}=-\nabla_i \nabla_j p
\end{equation}
Here $\bomega$ is the vorticity of the flow, $\bomega=\nabla \times {\bf v}$; .
$\rho_{ij}$ is the pressure hessian.
This equation is the direct consequence of the inviscid limit of the NSE and describes the
evolution of vorticity along  the trajectory of a liquid particle. It was for the first time obtained in \cite{Ohkitani-our_equation}.
The main assumption of the theory is that inside vortex filaments the 'longitudinal' (in relation to $\bomega$) part of the hessian doesn't depend on the local  vorticity, and is  determined by the large-scale component of a flow.$^1$ \footnotetext[1]{ An indirect confirmation of this statement  is presented in  \cite{Chevillard}, where the maximal eigenvalue of the pressure hessian is shown to be orthogonal to vorticity.}
This makes (\ref{main}) a linear stochastic equation in relation to $\bomega$. In \cite{JETP1}
 we wrote the corresponding equation for probability density function (PDF), $f(\bomega, \dot{\bomega}, t)$. Solving the equation and integrating over all angles and over $\dot{\omega}$, we found the intermediate asymptotic solution for the PDF at $\omega \to \infty$:
 $$ P(\omega) \propto 1 / \omega^4 $$
 In \cite{PRE2} we used it to derive the condition
 \begin{equation}
\label{cond}
 \min \limits _h (3h+2-D(h)) =0
\end{equation}
for the function $D(h)$.  As  follows from (\ref{MF}), this condition is just equivalent to
\begin{equation}
\label{zeta3}
\zeta_3 =1
\end{equation}
which is the well-known Kolmogorov's '4/5 law'.

\vspace{0.5cm}

The MF model is a dimensional theory, so it does not distinguish longitudinal and transverse scaling exponents. Also the condition (\ref{cond}) is valid for both of them. To describe the difference,  it is natural to use
two different functions $D(h)$ for $\zeta_n^{\perp}$ and $\zeta_n^{\parallel}$ \cite{Arn}.
But what is the relation between them,  why does the difference appear?

In \cite{PRL1, JETP2} we introduced a simple model of an axially symmetric vortex filament:
 \begin{equation} \label{vlocalhom}
v_r= a(t) r\,, \qquad v_z = b(t) z \,,\qquad v_{\phi}=\omega(t) r
\end{equation}
From the incompressibility condition and the Euler equation it then follows
\begin{equation}\label{a}
\begin{array}{l}
2a+b=0\,,\qquad \dot{a} +a^2 -\omega^2 = -P_1(t) \ ,
\\
\dot{\omega} +2 a \omega =0\,,\qquad \dot{b} +b^2 =-P_2(t) \ ,
\\  p =\frac12 P_1(t) r^2 + \frac12 P_2(t) z^2
\end{array}
\end{equation}
Taking the second derivative of $\omega$, we get
$$  
\ddot{\omega} = -P_2(t)\omega \ ,
$$ 
which corresponds to (\ref{main}). In accordance with the assumption discussed above, let $P_2(t)$ be a random function independent of $\omega$.
Then all moments of $\omega(t)$ increase exponentially as a function of time, $a\sim \dot{\omega} / \omega \ll \omega$ for large~$t$.

It is easy to calculate that in this model    $S^{\perp}_n \propto \langle \omega^n \rangle l^n $,
$S^{\parallel}_n \propto \langle a^n \rangle l^n $. In \cite{PRE} we discussed the connection between $t$
and $l$ and showed that $t \propto - \ln l$. Hence, in this model $S^{\perp}_n \gg S^{\parallel}_n$,
\ $\zeta_n^{\perp} < \zeta_n^{\parallel}$.

The equality of scaling exponents can be restored if we take  the other branch of the same filament
into account, assuming that the filament is
closed and 'turns back' at very large $|z|$. Then there are two
filaments with equal and opposite vorticities. The second branch
produces a perturbation of velocity $\delta v \sim \alpha \omega$,
where the coefficient $\alpha$ is inversely proportional to the
distance between the two branches.
(This is just analogous to an axial electric current producing magnetic field in
vacuum; one can write the corresponding solution to the Euler equation by
developing the perturbation as series in $\alpha$.)  The perturbation violates the
axial symmetry, so the scaling exponents become equal. However,
the pre-exponent of $S^{\parallel}_n$ is much smaller than that of
$S^{\perp}_n$ if $\alpha$ is small. The value of $\alpha$
increases as we approach the turnover point, and the two branches
become nearer. This gives us a hint that the regions where a
filament is strongly curved may make a small contribution to
$S^{\perp}_n$ but contribute much to $S^{\parallel}_n$.  This
effect is stronger for large $n$, since $S^{\parallel}_n\sim
\alpha^n S^{\perp}_n$. So, for infinitely large $n$ we can expect that
transverse structure functions are dominated by extremely stretched-out and very
thin, roughly axially symmetric filaments with very high vorticities, while for longitudinal 
structure functions one needs extremely curved parts of these filaments.

\vspace{0.5cm}

The discussed example is one of numerous solutions to the Euler equation that
are presented in a turbulent flow.
The observed exponents $\zeta_n^{\perp}$ and $\zeta_n^{\parallel}$  are produced by
contributions of many filaments. Different filaments (and different parts of them) make
contributions to different $n$ (or, in terms of MF theory, to
different $h$). This picture is very complicated, but, thanks to
the MF model, we don't need to know it in details. Approximating
$D(h)$ by a second-order polynomial \cite{PRE2} and making use of (\ref{cond}),
we only  have to fix one more point of the curve.

For this purpose, we consider the minimal possible $h=h_{\min}$.  This corresponds to $n\to \infty$.
 In \cite{PRE2} we have shown that $h_{\min}=0$: values $h<0$ are forbidden by the condition
 $d \zeta_n / d n \ge 0$ (see p.2).

 The existence of $h=0$ is proved by  the possibility of a  cylindric 
  filament with rotating velocity independent~of~$r$:
 \begin{equation}\label{extrim}
{\bf v} = [{\bf e}_z, {\bf r}/r]   
\end{equation}
 One can check that in this extreme case, indeed,  $\delta v (l) \sim l^0$. The velocity
 (\ref{extrim}) satisfies the Euler equation with pressure logarithmically divergent; this
 means that for any positive $h$, there exists a corresponding velocity distribution with
 converging pressure.

Calculation of the correlators directly from (\ref{extrim}) gives
under the limit $n\to \infty$:
\begin{equation}
\label{extrim2}
  \langle \left| \Delta v \times {\bf l}/l \right| ^n  \rangle \propto
  \frac{2^n}{n} \, l^2\,,\quad
  \langle \left| \Delta v \cdot {\bf l}/l \right| ^n \rangle \propto
   n^{-5/2} l^2
\end{equation}
The proportionality to $l^2$ is caused by the axial geometry of the filament (integrating $r dr$). This corresponds to the definition
of $D$ given in, e.g., \cite{Frisch}:
the probability that at least one of a pair of points would get inside the filament is proportional  to $l^2$. Thus, we get
  \begin{equation}
 \label{condperp}
D^{\perp}(0)=1
\end{equation}
To find the difference between $S^{\perp}_n$ and
$S^{\parallel}_n$,  we now remind that, at infinitely large $n$,
$S^{\perp}_n$ are contributed mostly by 'cylindric' parts of
filaments, while $S^{\parallel}_n$
 are dominated by point-like regions where  filaments are bent very strongly.
Actually, the behavior of the pre-exponents in (\ref{extrim2})
shows that the contribution of the 'extreme' filament to
$S_n^{\perp}$  increases as $n \to \infty$. To the opposite, its
contribution to $S_n^{\parallel}$ for large $n$ (i.e., small $h$)
is very small. There must be other solutions to the Euler equation
to determine the behavior of $D^{\parallel}(h)$ for small $h$ and,
equivalently, to make the most contribution to $S_n^{\parallel}$
for large $n$. These solutions correspond to 'strongly curved'
extreme filament. To satisfy $h=0$, velocity must be independent
of $r$. It may, for example, take the form
 ${\bf v} = (v_r(\theta), v_{\theta}(\theta), 0)$. 
  Such a solution  exists but it cannot be written analytically.
 At $\theta =0$, the radial velocity  diverges weakly:
$v_{\theta} \propto -\theta \sqrt{\ln\,(1/\theta)}\,, v_r \propto
\sqrt{\ln\,(1/\theta)}$; the pressure divergence is $p\propto \ln
r$, just as in (\ref{extrim}).   Since $\delta v \sim l^0$ in the
case, and averaging includes $r^2 dr$, the correlator is
proportional to $l^3$. This corresponds to
 \begin{equation}
 \label{condpar}
D^{\parallel}(0)=0
\end{equation}
The difference between the boundary values (\ref{condperp}) and
(\ref{condpar}) determines  the difference between the 
functions $D_{\perp}$ and $D^{\parallel}$.

\vspace{0.2cm}

We seek the solution $D(h)$ in the simplest form
$$
D(h) = 3- b(h-h_0)^2
$$
and use (\ref{cond}) and (\ref{condperp}),(\ref{condpar}) to find the two unknown parameters.
We obtain two equations:
$$
\begin{array}{ll}
 h^{\perp}_0 - \frac 38 (h^{\perp}_0)^2 = \frac 13 \ , \  b^{\perp}=2/(h^{\perp}_0)^2
\quad & \mbox{for} \ D^{\perp}
\\
 h^{\parallel}_0 - \frac 14 (h^{\parallel}_0)^2 = \frac 13 \ , \
 b^{ \parallel}=3/(h^{\parallel}_0)^2
\quad  & \mbox{for} \ D^{\parallel}
\end{array}
$$
Each of the equations has two roots. The bigger roots are
non-physical, because in this case the curve $\zeta_n$ would be
constant already at $n \ge n_*=D'_h(0)=2b h_0 \sim 2$,
and (\ref{cond}) would not hold. For the  second roots,                
in accordance with (\ref{MF}), we have
\begin{equation}   \label{result0}
\zeta_n = n h_0 - n^2 /4b  \ ,    
\qquad n \le n_* = 2 b h_0
\end{equation}
  Substituting the values of $h_0$,$b$,  we  get the scaling exponents:
\begin{equation}   \label{result}
\begin{array}{ll}
\zeta_n^{\parallel} = & \left\{
\begin{array}{lr}
 0.367 n - 1.12\cdot 10^{-2} n^2 \ , &   
\quad n \le  16.3  \ ; \\
3 \ , & \quad n>16.3 \ .
\end{array}     \right.
\\ 
\zeta_n^{\perp} = & \left\{
\begin{array}{lr}
 0.391 n - 1.91 \cdot 10^{-2} n^2 \ , &   
\quad n \le  10.2  \ ; \\
2 \ , & \quad n>10.2 \ .
\end{array}     \right.
\end{array}
\end{equation}
In Fig. 1 we compare this theoretical prediction with experimental
data by \cite{Benzi} and \cite{Gotoh}. We see that the result of our
simple model is very close to the experimental results and lies
inside the error bars of the experiments.

\begin{figure}
\vspace*{-0.4cm} \hspace*{-0.5cm}
\includegraphics[width=8cm]{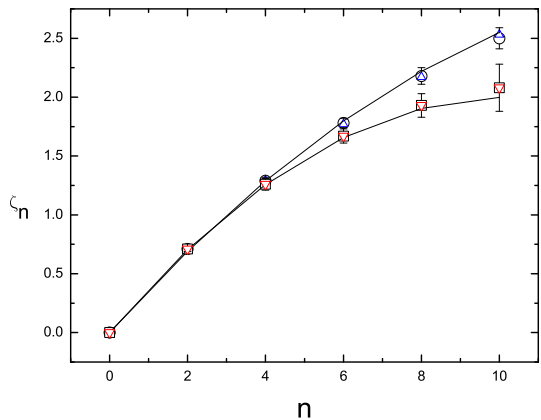}   
  \vspace*{-0.6cm}
\caption{Euler longitudinal (upper branch) and transverse (lower branch) scaling
exponents:  the results of DNS from \cite{Benzi} ($\bigcirc$,   $\square$)
and \cite{Gotoh}  ($\bigtriangleup$. $\bigtriangledown$), and the prediction of the theory
(\ref{result}) (lines). }
\end{figure}

Our simple model has one difficulty: the two parabolas plotted  in
Fig.~1 coincide at $n=0$ and $n=3$, hence they do not coincide at
$n=2$. This contradicts to the exact theoretical statement that
$\zeta_2 = \chi_2$. However, this difficulty is caused by
postulating the simplest parabolic shape for $D(h)$. It can easily
be solved by adding one more parameter and assuming  $D(h)$ to be
a cubic polynomial. Because of very small divergence ($1.6 \cdot 10^{-2}$) between $\zeta_2$ and $\chi_2$ in Fig.~1, the
coefficient by the eldest order would be very small ($\sim 10^{-4}$). It would
change very slightly (unnoticeable for an eye) the lines presented
in Fig.~1. The only thing they may  change significantly is the
rate of approaching the constant at large (but still intermediate)
$n$ (from 10 to 15, approximately). But this range of $n$ is,
anyway, badly described by the lowest-order polynomials: adding
more degrees of freedom with very small coefficients, though
unimportant for smaller $n$, would change the solutions for these
$n$.
However, the changes cannot be very big, since the exponents are still restricted by the values 
2 and 3, respectively.

One more comment is that, knowind $D(h)$, one can use the MF model to calculate, e.g., the PDF of velocity gradients or accelerations. In \cite{25years} it is shown that once $D(h)$ fits $\zeta_n$ well,  it would also
fit well the other quantities.

\vspace{0.3cm}

Thus, 
we propose an explanation of the difference between $\zeta_n^{\parallel}$
and $\zeta_n^{\perp}$ based on the difference between the filaments that contribute 
to the two structure functions: roughly speaking, this is the difference between 
axially symmetric and strongly curved ones. This allows to find the values of $\zeta_n$
 for very large $n$, and merging of the Vortex filament and Multifractal theories gives the
 whole functions.
The obtained solutions (\ref{result0})-(\ref{result})    
fit very well the observed scaling exponents   $\zeta_n^{\parallel}$
and $\zeta_n^{\perp}$, 
and we hope that they reveal the nature of
the difference between longitudinal and transverse structure
functions.

We thank Prof.A.V. Gurevich for his kind interest to our work. The work was partially supported by the RAS
Program 'Fundamental Problems of Nonlinear Dynamics'.

\end{document}